\documentstyle[12pt,a4]{article} 

\def\lsim{\mathrel{\lower2.5pt\vbox{\lineskip=0pt\baselineskip=0pt 
           \hbox{$<$}\hbox{$\sim$}}}} 
\def\gsim{\mathrel{\lower2.5pt\vbox{\lineskip=0pt\baselineskip=0pt 
           \hbox{$>$}\hbox{$\sim$}}}} 

\def\L{\Lambda}

\def\p{\partial}

\def\k{\kappa}
\def\a{\alpha}

\begin{document} 
\begin{flushright}
AUE-04-01\\ hep-th/0405281
\end{flushright}

\vspace{10mm}

\begin{center}
{\Large \bf 
Holographic dark energy model with non-minimal coupling}

\vspace{20mm}
 Masato Ito 
 \footnote{mito@auecc.aichi-edu.ac.jp}
\end{center}

\begin{center}
{
\it 
{}Department of Physics, Aichi University of Education, Kariya, 
448-8542, JAPAN
}
\end{center}

\vspace{25mm}

\begin{abstract}
We find that holographic dark energy model with non-minimally coupled scalar
field gives rise to an accelerating universe by choosing Hubble scale as IR cutoff.
We show viable range of a non-minimal coupling parameter in the
framework of this model.
\end{abstract}
\newpage 
\baselineskip=5.5mm

Recent astronomical observations indicate the facts that our universe
has an accelerated expansion and that cosmological constant is not zero
but extremely small and positive.
Thus we live in de Sitter universe.
From the point of view of particle physics, the origin of driving cosmic
acceleration must be explained, furthermore we would like to understand
reason why cosmological constant is positive.
The positive energy with negative pressure, dark energy, gives rise to
cosmic acceleration.
Several dark energy models including scalar fields, such as quintessence
or dilaton fields in string/M-theory, have been proposed.

Starting from a work of Fischler and Susskind based on holographic principle
\cite{'tHooft:gx,Susskind:1994vu},
the idea of holography has been introduced in cosmology 
\cite{Fischler:1998st}.
Motivated by Bekenstein entropy bound \cite{Bekenstein:ur,Bekenstein:ax}, 
the authors of \cite{Cohen:1998zx} have obtained the
current cosmological constant by suggesting a specific relationship
$L^{3}\L^{4}\lsim LM^{2}_{p}$, where
UV cutoff $\L$, IR cutoff $L$ and $M_{p}$ Planck scale. 
Furthermore it was argued that the holographic principle can provide a
natural solution to the cosmological constant problem 
\cite{Horava:2000tb,Thomas:2000km}.
This is because finite number of degrees of freedom within a region
prevents quantum correction to vacuum energy from diverging.
Remarkable paper of \cite{Li:2004rb} has proposed the holographic dark energy 
model with dark energy given by
\begin{eqnarray}
\rho_{\L}=\frac{3c^{2}}{8\pi G L^{2}}\label{eqn1}\,,
\end{eqnarray}
where $G$ is the gravitational constant and $c$ is the constant.
The holographic dark energy models are established by choosing an IR cutoff scale
such as Hubble scale, particle horizon and event horizon.  
The holographic dark energy model of Friedmann cosmology with the Hubble scale as IR
cutoff, in absence of scalar field, cannot give rise to an accelerating universe
due to dark energy without pressure \cite{Hsu:2004ri}.
On the other hand, papers of \cite{Li:2004rb,Huang:2004wt,Huang:2004ai} 
had argued that holographic dark energy models with
event horizon can produce an accelerating universe.
Moreover the holographic dark energy model had been applied to
Brans-Dicke theory as the extended version \cite{Gong:2004fq}.
The model indicated that the choice of Hubble scale or particle horizon
as IR cutoff is inconsistent with the conditions of accelerating universe.
While the choice of event horizon as the IR cutoff gives rise to an
accelerating universe with large Brans-Dicke parameter.
Furthermore the model must receive constraints of the time-varying
observations of $G$ 
because the time dependence of scalar field in Brans-Dicke theory brings about
the time variation of gravitational ``constant'' $\dot{G}/G$.
However the estimation of $\dot{G}/G$ isn't performed in \cite{Gong:2004fq}.
From our estimation, the model has acceptable value
$\dot{G}/G|_{0}=\alpha H_{0}\sim 10^{-14}{\rm yr}^{-1}$
by comparing with several observations 
\cite{Barrow:1996kc,Thorsett:1996fr,Uzan:2002vq,Benvenuto:2004bs}, 
where the parameter $\alpha$ shown in \cite{Gong:2004fq} is $\a\sim 10^{-3}$ and
subscript $0$ means the current value. 

In this letter we study whether the holographic dark energy model
with a non-minimally coupled scalar field
realizes an accelerating universe. 
We consider viable range of non-minimal coupling parameter $\xi$ from
restrictions of both Brans-Dicke parameter and time-variation of
gravitational constant. 
Theoretical restriction of $\xi$ had been
investigated by several authors 
\cite{Muta:mw,Mukaigawa:1997nh,Esposito-Farese:2000ij,Boisseau:2000pr}.

The action of the model considered here is given by \cite{Uzan:1999ch,Chiba:1999wt}
\begin{eqnarray}
S=\int\;d^{4}x\;\sqrt{-g}
\left(\frac{1}{2\k^{2}}{\cal R}-\frac{1}{2}\xi\phi^{2}{\cal R}
-\frac{1}{2}g^{\mu\nu}\p_{\mu}\phi\p_{\nu}\phi
\right)\label{eqn2}\,,
\end{eqnarray}
where $\k^{2}$ and $\phi$ are a bare gravitational constant and a scalar
field, respectively. 
The non-minimal coupling parameter between the scalar and the curvature is denoted
by $\xi$. 
Following ref.\cite{Gong:2004fq}, the action doesn't have particular
scalar potential.
However we must make some comments.
Although the possibility that vanishing scalar potential leads to singular
universe is pointed, the serious problem is beyond the scope of this paper.
In present paper we focus on the model of cosmic acceleration via
holographic dark energy, we are going to discuss the point elsewhere.

From Eq.(\ref{eqn2}) the effective gravitational constant can be expressed as
\begin{eqnarray}
8\pi G=\k^{2}\left(1-\xi\k^{2}\phi^{2}\right)^{-1}\label{eqn3}\,.
\end{eqnarray}
Here we take a flat four dimensional FRW metric
$ds^{2}=-dt^{2}+a^{2}(t)dx_{i}dx^{i}$, where $a$ is the scale factor.
We obtain field equations as follows \cite{Uzan:1999ch,Chiba:1999wt,Bertolami:1999dp}:
\begin{eqnarray}
&& H^{2}=\frac{\k^{2}}{3}\left(1-\xi\k^{2}\phi^{2}\right)^{-1}
\left(\rho+\rho_{\L}+\frac{1}{2}\dot{\phi}^{2}+6\xi H\phi\dot{\phi}\right)\label{eqn4}\,,\\
&&\ddot{\phi}+3H\dot{\phi}+6\xi\left(\dot{H}+2H^{2}\right)\phi=0\label{eqn5}\,,\\
&&\dot{\rho}+\dot{\rho_{\L}}+3H\left(\rho+\rho_{\L}+p+p_{\L}\right)=0\label{eqn6}\,,
\end{eqnarray}
where $H=\dot{a}/a$ is the Hubble parameter, $p$ and $\rho$ denote
pressure and density coming from background field, respectively.
The subscript $\L$ means component of dark energy.

We investigate a model of holographic dark energy with $c=1$ in
Eq.(\ref{eqn1}).
Taking the Hubble scale as IR cutoff in order to obtain value of
current cosmological constant \cite{Cohen:1998zx}, we have $L^{-1}=H$.
Consequently Eqs.(\ref{eqn1}) and (\ref{eqn3}) lead to the holographic dark
energy  
\begin{eqnarray}
\rho_{\L}=\frac{3}{\k^{2}}\left(1-\xi\k^{2}\phi^{2}\right)H^{2}\,.\label{eqn7}
\end{eqnarray} 
In the absence of scalar field, $\rho_{\L}$ is obviously the critical
energy density, as shown in ref \cite{Hsu:2004ri}, the conditions of
accelerated expansion are violated.
However, as mentioned later, non-minimally coupled scalar field
appearing in Eq.(\ref{eqn2}) allows existence of the solution
to the conditions of acceleration.
Also, the addition of a non-minimal coupling between the scalar and
the curvature had been justified in
quintessence model \cite{Uzan:1999ch,Chiba:1999wt,Bertolami:1999dp}. 

In order to obtain solutions to field equations in Eqs.
(\ref{eqn4})-(\ref{eqn6}),
we adopt power-law as specific solution.
Taking $a(t)=a_{0} t^{r}$ and $\phi(t)=\phi_{0}t^{s}$, 
Eq.(\ref{eqn5}) yields a relation between exponents $r$ and $s$:
\begin{eqnarray}
s(s-1)+3rs+6r(2r-1)\xi =0\label{eqn8}\,.
\end{eqnarray}
Moreover we consider the simplest case of dark energy dominated universe, 
namely, $\rho,p\ll 1$.
Therefore, Eq.(\ref{eqn4}) leads to the following relation:
\begin{eqnarray}
s=-12\xi r\label{eqn9}\,.
\end{eqnarray}
For $\xi=1/6$ corresponding to conformal coupling,
the solution satisfying Eqs.(\ref{eqn8}) and (\ref{eqn9}) is $r=s=0$.
The case isn't almost interesting because scale factor and scalar field
are constant.
We consider the case of $\xi\neq 1/6$.
Combining Eqs.(\ref{eqn8}) and (\ref{eqn9}), consequently, we obtain
\begin{eqnarray}
a(t)&=&a_{0}\;t^{\frac{1}{4(1-6\xi)}}\label{eqn10}\,,\\
\phi(t)&=&\phi_{0}\;t^{-\frac{3\xi}{1-6\xi}}\label{eqn11}\,.
\end{eqnarray}
Here $t=1$ means the present time.
Requiring accelerating universe $\ddot{a}>0$, we get 
inequality as follows:
\begin{eqnarray}
\xi >\frac{1}{8}\label{eqn12}\,.
\end{eqnarray}
In particular, for $1/8<\xi <1/6$, scale factor $a(t)$ and 
scalar field $\phi(t)$ are
increasing function and decreasing function, respectively.
For $\xi>1/6$, $a(t)$ and $\phi(t)$ are decreasing and increasing,
respectively.  
For your information, the case of $\xi <1/8$ corresponds to a
decelerating universe.
Below we perform more analyses from the range of $1/8<\xi <1/6$.

From Eqs.(\ref{eqn6}) and (\ref{eqn7}), the equation of state for holographic
dark energy is written as
\begin{eqnarray}
\omega_{\L}=\frac{p_{\L}}{\rho_{\L}}=
-\frac{1}{3}\left(24\k^{2}\xi^{2}\phi^{2}_{0}\left(t^{\frac{6\xi}{1-6\xi}}
-\xi\k^{2}\phi^{2}_{0}\right)^{-1}+48\xi-5\right)\label{eqn13}\,.
\end{eqnarray}
Accordingly we obtain the current $\omega_{\L}$ at $t=1$: 
\begin{eqnarray}
\omega_{\L 0}\simeq -\frac{48\xi-5}{3}\label{eqn14}\,,
\end{eqnarray}
where we assume that the current value of scalar field is extremely
small, $\xi\k^{2}\phi^{2}_{0}\ll 1$.
Restriction of $-1<\omega_{\L 0}<-2/3$ leads to the range of $\xi$:
\begin{eqnarray}
\frac{7}{48}<\xi<\frac{1}{6}\label{eqn15}\,.
\end{eqnarray}
From Eq.(\ref{eqn12}), note that the range of $\xi$ in
Eq.(\ref{eqn15}) is supported by current accelerating universe.

Moreover the present model must receive constraint via experiments of time
variation of gravitational constant. 
By performing differential of $G$ with respect to $t$, Eq.(\ref{eqn3})
leads to the value of current $\dot{G}/G$ under the assumption
$\xi\k^{2}\phi^{2}_{0}\ll 1$:
\begin{eqnarray}
\left.\frac{\dot{G}}{G}\right|_{0}\sim -24\xi^{2}\k^{2}\phi^{2}_{0}H_{0}
\lsim -\xi\k^{2}\phi^{2}_{0}\times 10^{-10}\;{\rm yr}^{-1}
\,,\label{eqn16}
\end{eqnarray}
where we used Eq.(\ref{eqn15}) and $H_{0}=(6-7)\times 10^{-11}{\rm yr}^{-1}$.
Thus positive $\xi$ implies that $G$ is a decreasing function with respect
to $t$.
Due to the assumption $\xi\k^{2}\phi^{2}_{0}\ll 1$, $\dot{G}/G$ of the present model
isn't excluded by observations
\cite{Barrow:1996kc,Thorsett:1996fr,Uzan:2002vq,Benvenuto:2004bs}.
According to ref.\cite{Chiba:1999wt}, the Brans-Dicke parameter
is expressed as $\omega_{\rm BD}\sim (\xi\k^{2}\phi^{2}_{0})^{-1}\gg 1$ 
under the assumption.
Thus, in the present model, the non-minimal coupling parameter of
limited range ($0.146\lsim \xi\lsim 0.167$) provides an accelerating universe
to be consistent with observations. 

In contrast to the above assumption, we consider $\xi\k^{2}\phi^{2}_{0}\gg 1$.
Eq.(\ref{eqn13}) leads to $\omega_{\L 0}\simeq -(24\xi -5)/3$,
and the requirement of $-1<\omega_{\L 0}<-2/3$ yields
$7/24<\xi<1/3$ which is consistent with the condition of acceleration.
Furthermore the time variation of $G$ is given by $\dot{G}/G|_{0}\sim 24\xi
H_{0}\lsim 10^{-10}{\rm yr}^{-1}$ which is allowable value if compared
with the results of observations \cite{Barrow:1996kc,Thorsett:1996fr,
Uzan:2002vq,Benvenuto:2004bs}.
On the other hand the Brans-Dicke parameter of this case is given by 
$\omega_{\rm BD}\sim-1/4\xi$, namely this situation obviously conflicts with current
$\omega_{\rm BD}>500$.
Consequently it turns out that the case of $\xi\k^{2}\phi^{2}_{0}\gg 1$
is ruled out.
Since effective gravitational constant in Eq.(\ref{eqn3})
becomes negative, anti-gravity is explicitly excluded.

In conclusion, we pointed out a possibility that the holographic
dark energy model with a non-minimally coupled scalar field gives rise to an
accelerating universe.
Ref \cite{Gong:2004fq} indicated that the holographic dark energy applied
to Brans-Dicke theory cannot give rise to an accelerating universe by
choosing Hubble scale as IR cutoff. 
However, despite of a choice of Hubble scale, it was shown that the
Einstein-Hilbert action plus a non-minimal coupling between scalar and
curvature can lead to cosmic acceleration.
The viable range of the non-minimal coupling parameter to be consistent
with the current observations is shown.
Thus the holographic dark energy model can provide a new scenario of
cosmological model.
%
%

\end{document}